**Vlasov Simulation of Emissive Plasma Sheath with Energy-Dependent Secondary Emission Coefficient and Improved Modeling for Dielectric Charging Effects**

Guang-Yu Sun[1,2], Shu Zhang[1], Bao-Hong Guo[3], An-Bang Sun[1*], Guan-Jun Zhang[1*]

[1]State Key Laboratory of Electrical Insulation and Power Equipment, School of Electrical Engineering, Xi'an Jiaotong University, Xi'an 710049, People's Republic of China

[2]Ecole Polytechnique Fédérale de Lausanne (EPFL), Swiss Plasma Center (SPC), CH-1015 Lausanne, Switzerland

[3]Centrum Wiskunde & Informatica (CWI), Amsterdam, The Netherlands

Corresponding Author
anbang.sun@xjtu.edu.cn
gjzhang@xjtu.edu.cn

**Abstract**

A one-dimensional Vlasov-Poisson simulation code is employed to investigate the plasma sheath considering electron-induced secondary electron emission (SEE) and backscattering. The SEE coefficient is commonly treated as constant in a range of plasma simulations, here improved SEE model of a charged dielectric wall is constructed which includes the wall charging effect on SEE coefficient and the energy dependency of SEE coefficient. Pertinent algorithms to implement above SEE model in plasma simulation are studied in detail. It is found that the SEE coefficient increases with the amount of negative wall charges, which in turn reduces the emissive sheath potential. With energy-dependent SEE coefficient, the sheath potential is a nonlinear function of the plasma electron temperature, as opposed to the linear relation predicted by classic emissive sheath theory. Simulation combining both wall charging effect and SEE coefficient' energy dependency suggests that the space-charged limited sheath is formed at high plasma electron temperature levels, where both sheath potential and surface charging saturate. Additionally, different algorithms to implement the backscattering in kinetic simulation are tested and compared. Converting backscattered electron to secondary electron via an effective SEE coefficient barely affects the sheath properties. The simulation results are shown to be commensurate with the upgraded sheath theory predictions.

## 1 Introduction

Plasma sheath is a non-neutral space charge region which appears between bulk plasma and plasma-facing components. A sheath becomes emissive due to surface emission processes, including secondary electron emission (SEE), backscattering, field emission, thermionic emission, photoemission, etc. Emissive sheath widely appears in confined laboratory plasmas and plays a vital role in numerous industrial plasma applications such as plasma processing, electric proportion, plasma diagnostics, plasma source and many others (1-4). The present research focuses on the algorithms to implement the interactions between plasma and dielectric surface in the kinetic simulation and the underlying sheath physics.

The classic emissive sheath theory was first established by Hobbs and Wesson by analyzing the current balance near a floating emissive boundary (5). It was proved that the presheath structure is not significantly affected by boundary emission and the emissive sheath potential $\varphi_{sh}$ is a function of the surface emission coefficient defined as $\gamma_e = \frac{\Gamma_{em}}{\Gamma_{ep}}$, with $\Gamma_{em}$ the surface emission flux and $\Gamma_{ep}$ the incoming plasma electron flux:

$$e\varphi_{sh} = T_e \ln[(1-\gamma_e)\sqrt{\frac{\mu}{2\pi}}] \qquad (1)$$

Here $T_e$ is the plasma electron temperature and $\mu = m_i/m_e$ is the ion-electron mass ratio. The study of emissive sheath has been persistently developed since then, and a range of emissive sheath theories have been proposed (6-9). For strongly emissive surface, a space-charge limited (SCL) sheath with nonmonotonic potential profile is formed when $\gamma_e$ exceeds certain critical value (10). More recent study suggested that the SCL sheath cannot remain stable if cold ions are generated by charge exchange collisions in sheath (11). The emissive sheath is difficult to be directly accessed by conventional probe diagnostics due to its small size, so sheath theories are frequently validated against numerical simulations (12-14).

Common simulation approaches of plasma-surface interaction consist of particle model, fluid model, kinetic model, global model, etc. Particle model provides self-consistent plasma dynamics based on first principles, which usually requires large quantities of computational resources. Fluid simulation is usually cheaper in terms of numerical cost, whereas one issue that potentially hinders the simulation accuracy is whether the fluid assumptions remain valid in the sheath region, particularly with non-thermal plasmas. Consequently, the sheath itself is frequently taken as a boundary condition instead of being simulated in the fluid model. The kinetic simulation captures the physics that are neglected in fluid model when averaging over the velocity moments, but exhibits lower performance when simulating complex reactions. In addition, the use of collision operators is inevitably less precise compared to the particle model. Yet one advantage of the kinetic simulation model is that it evades the statistical noise that must be reduced by using large macroparticle number in the particle model, and provides smooth profiles for every time step. This benefit is particularly obvious for the nonlinear plasma behaviors in the sheath study. In addition, it is easier to directly control certain physical quantities to facilitate comparison with theory prediction in a kinetic model. This is why numerous kinetic models are employed in the study of sheath-related topics (15-18).

Boundary electron emission due to SEE is widely implemented in the numerical modeling of confined plasma. The SEE coefficient $\gamma_e$ depends on the incident electron energy and direction. SEE is hence intrinsically coupled with the plasma properties (19). Additionally, the SEE coefficient is linked with the cumulative plasma fluxes, as the wall charges affect the extraction of excited internal secondary electrons (SEs) inside the material surface. Proper algorithms are therefore needed in order to implement these effects in the numerical simulation, which is the focus of the present work.

Paper is structured as follows: section 2 introduces the employed simulation model, section 3 studies the methods to configure appropriate boundary condition considering dielectric wall charges and the energy dependency of SEE coefficient. The simulation setup involving electron backscattering is also expatiated. All simulation results are validated against upgraded emissive sheath theories and the discrepancies between theory and simulation are analyzed.

## 2  Simulation setup

In this section, employed simulation algorithms and typical simulation results are presented. The simulation model is inspired by previous Vlasov-Poisson solvers (11, 15), and a basic version of the code was used in our recent simulations (13, 20, 21). The employed 1D1V kinetic simulation code is based on the following kinetic equation:

$$\frac{\partial f_s(x,v_s)}{\partial t} + v_s \frac{\partial f_s(x,v_s)}{\partial x} + \frac{q_s E(x)}{m_s}\frac{\partial f_s(x,v_s)}{\partial v_s} = \frac{\partial f_s(x,v_s)}{\partial t}\bigg|_{coll} \quad (2)$$

Here the subscript $s$ represents species, $m_s$, $q_s$, $v_s$ are mass, charge and velocity of the species, $E$ is the electric field and the RHS is the collision-source term. The code simulates the evolution of velocity distribution functions (VDFs) according to Equation (2). In the present work, only electrons and singly charged ions are considered. The electron and ion velocity distribution functions are initialized as uniformly distributed Maxwellian with temperatures $T_e$, $T_i$ and density $n_0$:

$$f_{s0}(x,v_s) = n_0 \sqrt{\frac{m_s}{2\pi T_s}} exp(-\frac{m_s v_s^2}{2T_s}), \quad (3)$$

For the study of sheath properties, keeping a constant bulk plasma density facilitates the parameter scan. The collision term is separated into two parts, one plasma source term which compensates for the boundary particle losses, and one relaxation term which restores the plasma VDFs back to equilibrium at a constant rate (the BGK collision operator).

$$\frac{\partial f_s(x,v_s)}{\partial t}\bigg|_{coll} = S_{source} + v_s [\frac{n(x)}{n_0} f_{s0} - f_s(x,v_s)] \quad (4)$$

The collision frequency $v_s$ is set as being proportional to the inverse of particle transit time, i.e. $v_e \propto v_{eTh}/L$ and $v_i \propto v_{cs}/L$ with $v_{eTh} = \sqrt{\frac{T_e}{m_e}}$ the electron thermal speed, $v_{cs} = \sqrt{\frac{T_e}{m_i}}$ the sonic speed and $L$ the spatial range of simulation domain. In the present simulation, $v_e = 10 v_{eTh}/L$ and $v_i = 5 v_{cs}/L$ are adopted. The coefficients are chosen to guarantee that the bulk plasma VDFs are not strongly diverged from the equilibrium. The charge source term is uniformly distributed in the region $(0.1L, 0.9L)$, which equals to $S_{source} = 1.25 \frac{\Gamma_i}{n_0 L} f_{s0}$. $\Gamma_i$ is total ion flux at two boundaries. The ion BGK relaxation is also turned on only in the region $(0.1L, 0.9L)$. These two treatments avoid cold ion generation near the surface, such that a SCL sheath is not destroyed to form an inverse sheath. The present work mainly focuses on classic Debye sheath. Note that the ionization term here is simplified and does not consider realistic ionization collision, which aims at fixing the bulk plasma density at the desired level. More physical collision operator was adopted in previous numerical modeling, (22) where the ionization source is calculated by an integral of the ionization cross section, velocity and background neutral density over EVDF. Such treatment provides more self-consistent simulation results where ionization rate is closely coupled with the local EVDF. Implementing such realistic collision operator in the present work will to some extent alter the bulk plasma properties but the general validity of obtained conclusions should be intact.

Boundary conditions of VDFs are critical for the implementation of abovementioned surface processes. For a surface with only secondary electron emission, the emitted secondary electrons are assumed to follow half-Maxwellian distribution with temperature $T_{em}$. Taking the left boundary as an example, the EVDF boundary condition is $f_{em} = f_e(v_e)|_{x=0, v_e > 0} = n_{em} \sqrt{\frac{2 m_e}{\pi T_{em}}} \exp(-\frac{m_e v_e^2}{2 T_{em}}) = C \exp(-\frac{m_e v_e^2}{2 T_{em}})$, $n_{em}$ is the emitted electron density at the boundary. The factor $C$ is determined by the definition of SEE coefficient, such that the emitted electron flux satisfies $\Gamma_{em} = \int_0^\infty C \exp(-\frac{m_e v_e^2}{2 T_{em}}) v_e dv_e = \gamma_e \Gamma_{ep}$, with plasma electron flux $\Gamma_{ep} = \left| \int_{-\infty}^0 f_e(v_e)|_{x=0} dv_e \right|$ for the left boundary. Taking absolute value means that flux is by default positive. Above assumptions yield $f_{em} = \frac{m_e}{T_{em}} \gamma_e \Gamma_{ep} \exp(-\frac{m_e v_e^2}{2 T_{em}})$. This expression has been used in previous studies using basic version of the adopted simulation code (20). The boundary condition will be further reviewed in section 3 with improvements considering energy dependency and charging effects of the factor $\gamma_e$. Since no ion reflection is considered, IVDF boundary condition is simply $f_i(v_i)|_{x=0, v_i > 0} = 0$ for the left boundary. The right boundary is symmetrical to the left in the present work.

A flow chart is given in Figure 1 to visualize the simulation procedure. For each time step, two advections are performed with explicit finite difference method in upwind scheme, then the source term and BGK collision operator are applied. The velocity advection requires the electric field distribution, which is solved by assuming zero electric field in the center of simulation domain, and then integrating the net charge density towards the boundary using Gauss's Law. The simulation usually converges in several μs with time step of 0.01ns. The simulation convergence at $t = t_{end}$ is assumed to be achieved when parameters including sheath potential, wall charge density, plasma flux at wall all have small variations with discrepancy lower than 0.1% in $10^4$ time steps. The time step is dictated by the Courant–Friedrichs–Lewy (CFL) stability criterion.

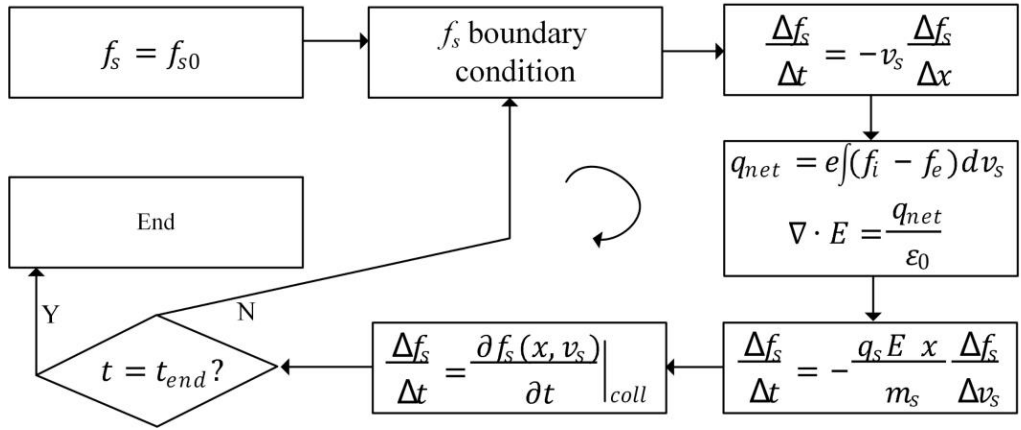

Figure 1. Schematic of the program execution.

Other choices of simulation parameter are introduced below. Plasma density is $5\times10^{14}$ m$^{-3}$, plasma electron temperature is 10eV by default, ion temperature is 0.1eV, emitted electron temperature is 2eV, scale of simulation domain is 10cm, spatial resolution is $10^{-4}$m, electron and ion velocity ranges cover 8 times the electron thermal speed and sound speed, divided into $10^3$ points. Typical simulation results including potential, density and velocity distribution function with above settings are shown in Figure 2.

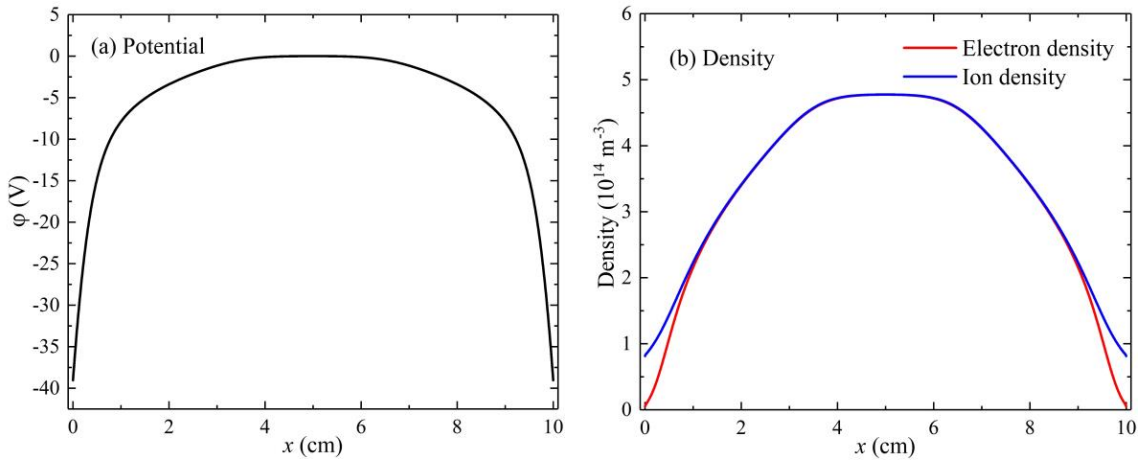

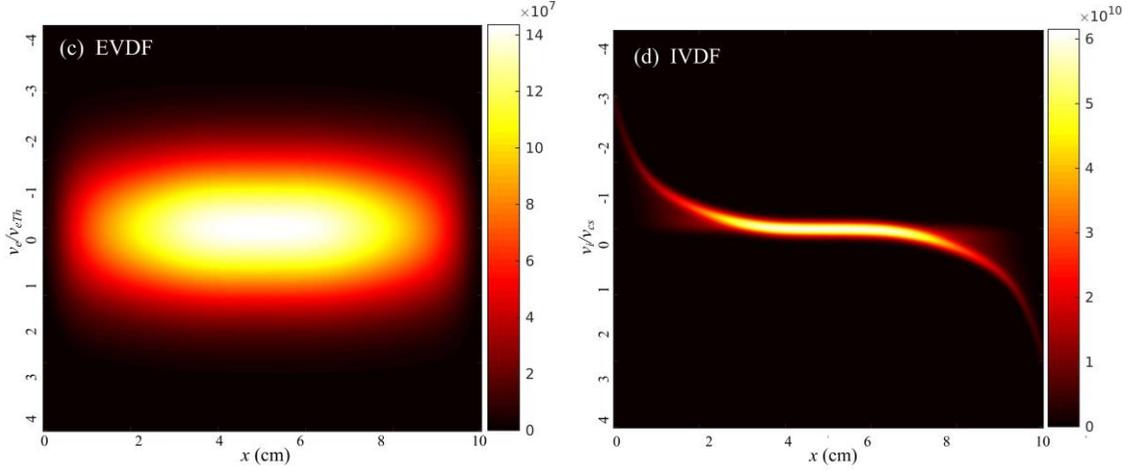

Figure 2. Typical simulation results obtained by the 1D1V kinetic simulation. (a) Space potential. (b) Electron and ion density. (c) Electron velocity distribution function. (d) Ion velocity distribution function.

## 3 Simulation results and model validation against theory
### 3.1 Influence of charge trapping on SEE coefficient

A primary electron (PE) penetrating into material surface, without being backscattered, generates internal secondary electrons while being slowed down. Only a fraction of the internal SEs are transported to the surface and escape from the material, becoming true SEs. The transport of internal SEs is characterized by the escape mean free path $\lambda_{es}$, with the escape probability $P_{es} = \exp(-\frac{x}{\lambda_{es}})$. The position $x$ is counted from where a SE is generated towards the material surface. Integrating $P_{es}$ multiplied by the generation function $g_{SE}$ over the whole range of PE's trajectory, aka the primary range $R_p$, yields the SEE coefficient:(23)

$$\gamma_{e0} = \int_0^{R_p} g_{SE}(x, \varepsilon_{pe}) P_{es}(x) dx \tag{5}$$

The term $\gamma_{e0}$ represents the initial SEE coefficient for an uncharged wall when considering the charging effect, to be distinguished from $\gamma_e$ for a charged wall. By convention, the generation function is replaced by $1/\lambda_{SE}$ to facilitate derivation. The parameter $\lambda_{SE}$ is the mean free path of SE's generation and has dimension of meter. Equation (5) is calculated to be:

$$\gamma_{e0} = \frac{\lambda_{es}}{\lambda_{SE}}[1 - \exp(-\frac{R_p}{\lambda_{es}})] \approx \frac{\lambda_{es0}}{\lambda_{SE}}, \tag{6}$$

supposing $R_p \gg \lambda_{es}$.

Equation (6) indicates that SEE is dictated by two characteristic lengths: $\gamma_{e0} = 1$ if the mean free path to create a SE equals to its escape mean free path, $\gamma_{e0} < 1$ if SEs vanish quicker than their creation during the transport to the surface, vice versa.

The value of $\lambda_{SE}$ is in general independent of the surface charging, and the surface charging mainly affects $\gamma_{e0}$ via the escape mean free path of SE, as internal SEs are captured by trap states or recombine with holes. These processes are prescribed by the density of trap state and hole as well as the corresponding cross-sections:

$$\lambda_{es0}^{-1} = N_T \sigma_T + N_h \sigma_R, \tag{7}$$

Here $N_T$, $N_h$ are intrinsic densities of trap states and holes, and $\sigma_T$ and $\sigma_R$ are cross-sections for trapping and recombination. For a classic Debye sheath, the plasma-facing wall is negatively charged, which means that some electron trap states are occupied. The escape mean free path of SE in a dielectric with trapped electron density $n_T$ (dimension m$^{-3}$) is therefore expressed as:

$$\lambda_{es}^{-1} = (N_T - n_T)\sigma_T + N_h \sigma_R, \tag{8}$$

with $n_T \leq N_T$. Combining Equations (6)-(8), the SEE coefficient with trapped charge is:

$$\gamma_e = \frac{\gamma_{e0}}{1-\lambda_{es0}n_T\sigma_T} \tag{9}$$

Equation (9) suggests that the dielectric surface immersed in a plasma becomes more emissive (larger $\gamma_e$) as negative surface charges accumulate. Note that this trend will not develop without limit, as higher surface emissivity decreases the amount of trapped charges, which in turn decreases $\gamma_e$. The increase of $\gamma_e$ also halts at the limit of $n_T = N_T$, which is however unlikely to be achieved for most laboratory plasmas. The sheath stability issue considering charge trapping will be addressed later in section 3.3.

The term $n_T$ however warrants more discussions before implementing surface charging effects in plasma simulation. The trapped charge density, to be distinguished from plasma density, represents the charges located in a layer much thinner than the plasma sheath. The surface layer of dielectric material, where trapped charges reside, is usually studied in nanometer scale (24). In the present simulation, the spatial grid resolution is 100μm, which is well below the limit set by the Courant–Friedrichs–Lewy condition but is still above nanometer scale by several orders of magnitude. It is hence difficult to simulate plasma coupled with dielectric surface layer in real time. Therefore, Equation (9) is reformed as follows supposing all trapped charges are closely attached to the surface and instantaneous charge transport process in surface layer is assumed:

$$\gamma_e = \frac{\gamma_{e0}}{1-K_{wall}|\sigma_{wall}|} \tag{10}$$

The wall charging factor $K_{wall}$ is expressed as $K_{wall} = \lambda_{es0}\sigma_T/(ed_{sl})$, $\sigma_{wall}$ is the wall charge density (Cm$^{-2}$), $d_{sl}$ is the depth of dielectric surface layer. The expression $n_T = |\sigma_{wall}|/ed_{sl}$ is used when deriving Equation (10). The updated $\gamma_e$ expression allows for facile implementation of the surface charging effects in simulation models. The wall charge density is calculated by monitoring the surface electron and ion flux at each time step. The factor $K_{wall}$ is adjustable which contains all the necessary information of the charge transport process in dielectric surface layer. $\gamma_{e0}$ is the uncharged SEE coefficient assigned in the beginning of simulation.

A critical issue is then to determine the order of magnitude of the factor $K_{wall}$ for general applications of the above theories in plasma simulations. A rough estimation is given below based on classic SEE theories for solid material. The most uncertain factor is the trapping cross-section $\sigma_T$ as it varies strongly with the dielectric material and accurate theory prediction is in general difficult. From a range of experimental measurements (25-28), the trapping cross-section is between $10^{-13}$-$10^{-11}$cm$^2$ but lower values exist. Here to test the maximum effects of surface charging, the upper limit $10^{-11}$cm$^2$ is taken. $\lambda_{es0}$ is calculated from $\lambda_{es0} = \lambda_{SE}\gamma_{e0}$, with $\lambda_{SE}$ expressed by $\lambda_{SE}^{-1} = C\frac{1}{\varepsilon_{ion}}\left|\frac{d\varepsilon_{pe}(x)}{dx}\right|$. Here $C \leq 1$ and $\varepsilon_{ion}$ is the mean excitation energy for one SE, (29) approximated by $\varepsilon_{ion} = 3\varepsilon_g + 1eV$ (27). $\varepsilon_g$ the gap energy of dielectric material. $\frac{d\varepsilon_{pe}(x)}{dx}$ is expressed by an empirical formula depending on the primary range and the primary electron current gradient (27). Our calculation suggests that $\lambda_{es0}/d_{sl}$ is within $10^1$-$10^2$, which eventually gives $K_{wall}$ of $10^5$-$10^6$ C$^{-1}$m$^2$. Above estimation is inevitably subject to some levels of arbitrariness, and smaller values are possible since a peak $\sigma_T$ value is chosen here.

Below a range of $\gamma_{e0}$ values are assigned as initial conditions in the kinetic simulation, with different wall charging factors $K_{wall}$, to study the effects of surface charging on emissive sheath properties. Other plasma parameters are kept constant. The charge trapping increases $\gamma_e$, which in turn reduces sheath potential and amount of wall charges, shown in Figure 3(a)-(c). Note that the charge conservation of a floating boundary requires that the amount of negative charges in dielectric wall should be equal to the net positive charges in sheath, hence the sheath potential and wall charge density behave collectively. The value of $\gamma_{e,final}$ ($\gamma_e$ value of a converged simulation run) scales up with $K_{wall}$, and cases with $K_{wall} = 4E6$ C$^{-1}$m$^2$ exhibit a 35.5% to 87.2% improvement relative to the cases without charging effect in the selected range of $\gamma_{e0}$.

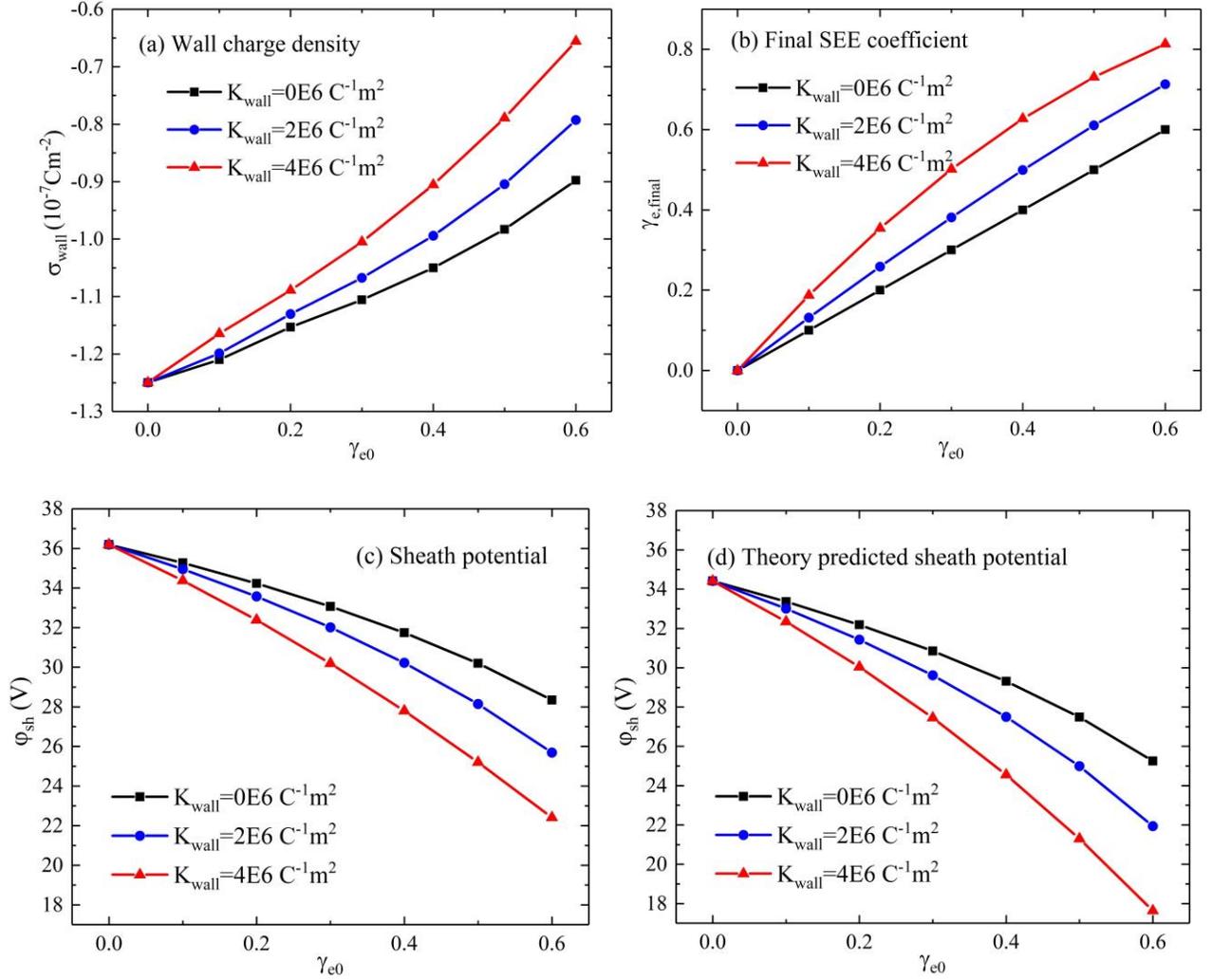

Figure 3. Emissive sheath properties with SEE considering wall charging effect, for a range of $\gamma_{e0}$ with different wall charging factors. (a) Wall charge density. (b) Final SEE coefficient after reaching convergence. (c) Sheath potential. (d) Sheath potential predicted by emissive sheath theory.

Figure 3(d) shows the theoretical prediction of sheath potential according to Equation (1) and Equation (10). Particularly, the black curve in Figure 3(d) is exactly the emissive sheath potential predicted by Hobbs, serving as a benchmark for the code validity. Since the sheath potential in the simulation is counted from the wall to the central plasma, it includes both the sheath and presheath potential drop. A better way of presentation is to determine the presheath location from simulation and subtract it from the simulated total potential drop, but the determination of presheath incurs some uncertainties when analyzing simulation results, so the total potential drop is considered in the present work. The calculated emissive sheath potential from Equation (1) is hence augmented by $e^{0.5}T_e$ in Figure 3(d), based on the Bohm presheath criterion. Same treatment is applied in the following sheath potential calculations. Note that the Bohm presheath was proved not to be affected by the SEE (30). It is clear that the simulation results and theory predictions agree considerably well. The discrepancies between Figure 3(c) and (d) are small when charging effects are not considered ($K_{wall}=0$), with a difference of 4.9-10.1% for the considered range of $\gamma_{e0}$. For cases considering charging effects, the peak discrepancy becomes 14.6% and 21.3%, for $K_{wall}=2E6 \; C^{-1}m^2$ and $K_{wall}=4E6 \; C^{-1}m^2$, respectively. The discrepancies consist of the intrinsic discrepancy between simulation and theory without charging effect, and the inconsistent choice of charge density when calculating $\gamma_e(n_T)$. The intrinsic discrepancy is likely related to the limited presheath region considered in the simulation. The discrepancy due to charging effect is because the wall charge density should be self-consistently determined by the sheath potential, whereas the exact spatial potential profile is unknown from the emissive sheath theory. Since the theory-predicted sheath potentials are smaller than the simulated results without charging effects, using larger wall charge densities from simulation (due to larger sheath potential) further increases the value of $\gamma_e$ and decreases the calculated sheath potential, leading to larger discrepancy in the end.

## 3.2 Energy dependency of SEE coefficient

In section 3.1, the SEE coefficient is constant for all incident electron energies. This is convenient for comparison with the classic emissive sheath theory (Equation (1)) by Hobbs and Wesson (5), but is oversimplified as $\gamma_{e0}$ has a strong dependency on the incident electron energy $\varepsilon_{PE}$ and angle $\theta_{PE}$. In the present 1D1V simulation, normal incidence is assumed. The SEE coefficient usually first increases with the primary electron energy up to a threshold energy level $\varepsilon_{max}$ with peak value $\gamma_{max}$, then begins to decrease. Note that $\varepsilon_{max}$ is in general several hundred eV and is therefore well above $T_e$ for most industrial plasmas and even fusion plasma in scrape-off layer (SOL) near plasma-facing components. A typical SEE coefficient curve is shown in Figure 4, where primary electron energy and dimensionless SEE coefficient $\gamma_{e0}$ are normalized by two coefficients $\varepsilon_{max}$ and $\gamma_{max}$. The following empirical formula is employed to derive the SEE coefficient curve: (31)

$$\gamma_{e0}(\varepsilon_{PE}, \theta_{PE}) = 1.526\gamma_{max}\left(1 + \frac{k_s\theta_{PE}^2}{2\pi}\right)[1 - \exp(-z^{1.725})]/z^{0.725} \quad (11)$$

$$z = 1.284\varepsilon_{PE}/[\varepsilon_{max}(1 + \frac{k_s\theta_{PE}^2}{\pi})]$$

(12) where smoothness factor $k_s = 1$ and incident angle $\theta_{PE} = 0$ are applied and $z$ is also a dimensionless factor. In the energy range of industrial plasma electron, e.g. below 100eV, a linear approximation provides good estimate of $\gamma_{e0}$, with $\gamma_{e0,estimate}(\varepsilon) = \frac{\varepsilon}{\varepsilon_1}$. Here $\varepsilon_1$ the required primary electron energy that produces one secondary electron. The values of $\varepsilon_1$ are determined by setting $\gamma_{e0} = 1$ in Equation (11). A comparison of linear approximation and real $\gamma_{e0}$ is shown in the subplot of Figure 4. A range of such empirical formulae exist which give similar profiles of $\gamma_{e0}$ (19, 32).

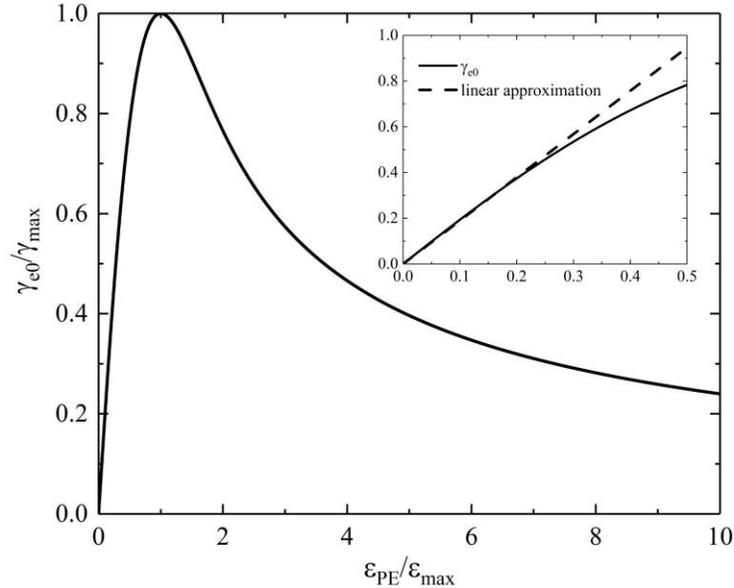

Figure 4. SEE coefficient as a function of incident electron energy. Normalization over $\varepsilon_{max}$ and $\gamma_{max}$ is applied. The subplot shows the comparison of $\gamma_{e0}$ with its linear approximation in low energy range.

In boundary plasma simulations, the SEE coefficient is applied in a variety of ways depending on the employed simulation approach. For particle-in-cell (PIC) simulation, the implementation is straightforward, as $\gamma_{e0}$ of each super-particle is calculated individually by Equation (11), (12). The decimal part of calculated $\gamma_{e0}$ is usually saved for the following super-particles until it cumulates up to one. For fluid simulation, the whole sheath region is commonly characterized by a boundary condition, e.g. the sheath heat transmission coefficient which constitutes the boundary condition of electric potential and heat flux, is sensitive to $\gamma_{e0}$. Since the sheath region is usually not simulated, the $\gamma_{e0}$ is calculated by $\gamma_{e0}(\langle\varepsilon_{e,wall}\rangle)$, with $\langle\varepsilon_{e,wall}\rangle$ the estimated mean electron incident energy at wall. For kinetic simulation, $\gamma_{e0}$ is better calculated by averaging $\gamma_{e0}$ over the EVDF at wall $f_{e,wall}$. Three different methods to apply SEE in the present simulation are tested, namely 1) $\gamma_{e0} = \gamma_{e0}(T_e/2)$ and $\Gamma_{em} = \gamma_{e0}\Gamma_{ep}$, with $T_e/2$ obtained by integrating electron kinetic energy over Maxwellian EVDF, 2) $\gamma_{e0} = \gamma_{e0}(\langle\varepsilon_{e,wall}\rangle)$ and $\Gamma_{em} = \gamma_{e0}\Gamma_{ep}$, with $\langle\varepsilon_{e,wall}\rangle = \int_0^\infty 0.5m_e v_e^2 f_{pe,wall} dv / \int_0^\infty f_{e,wall} dv$, and 3) $\Gamma_{em} = \int_0^\infty \gamma_{e0}(0.5m_e v_e^2) v_e f_{pe,wall} dv$, with equivalent SEE coefficient $\gamma_{e0} = \Gamma_{em}/\Gamma_{ep}$, $\Gamma_{ep} = \int_0^\infty v_e f_{pe,wall} dv$. $\gamma_{e0}$ is updated at each time step for the last two methods. Scans of assigned plasma electron temperature in simulation $T_e$ are performed with two different ($\varepsilon_{max}$, $\gamma_{max}$) sets resembling typical dielectric wall materials, for the three approaches to implement SEE coefficient. The obtained SEE coefficient and sheath potential are shown in Figure 5.

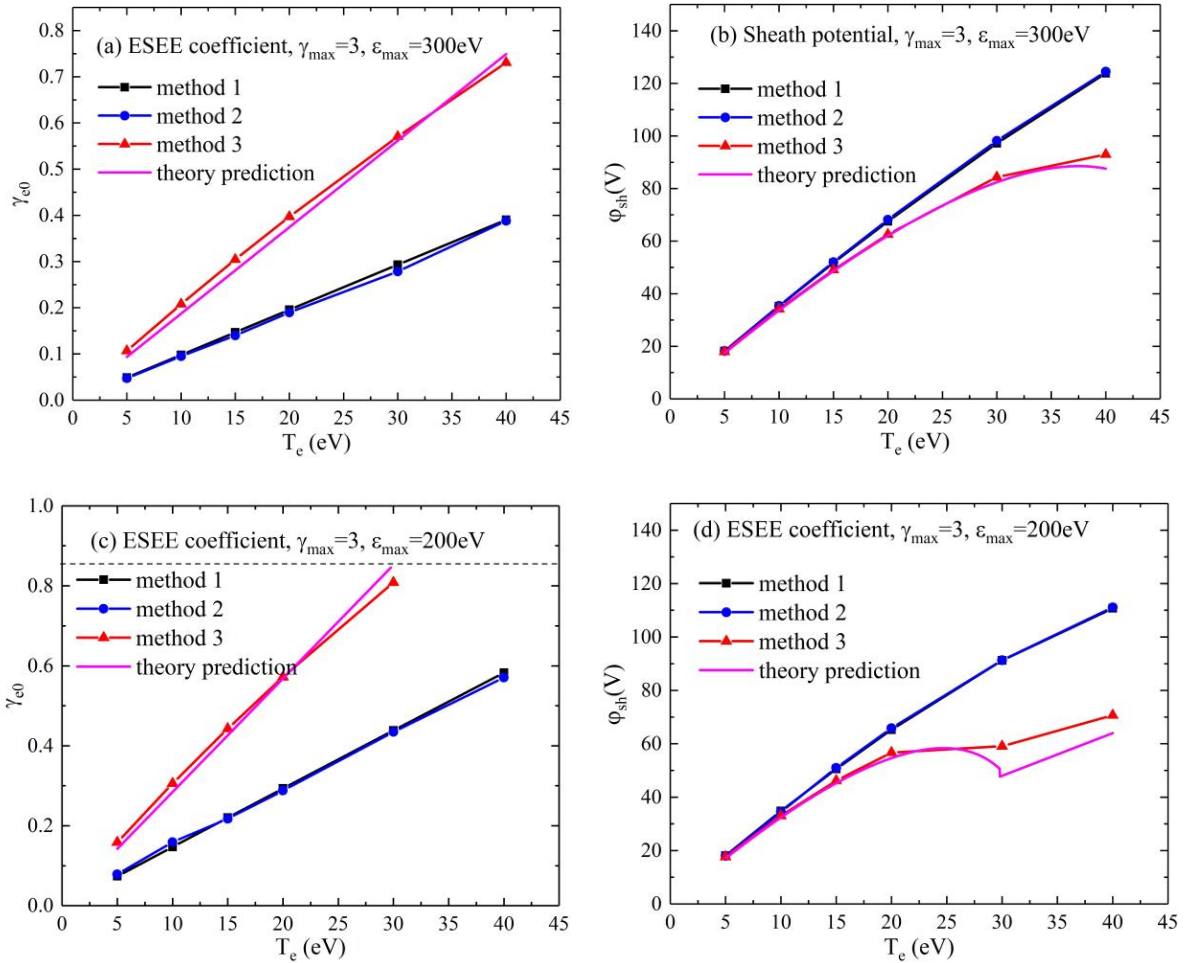

Figure 5. SEE coefficient and sheath potential calculated with 3 different $\gamma_{e0}$ setup methods obtained by simulation, in addition to the theory prediction. A series of plasma electron temperature and two different ($\varepsilon_{max}$, $\gamma_{max}$) sets are applied. (a)-(b) $\varepsilon_{max} = 300eV$, $\gamma_{max} = 3$, (c)-(d) $\varepsilon_{max} = 200eV$, $\gamma_{max} = 3$, (a), (c) show $\gamma_{e0}$ and (b), (d) show $\varphi_{sh}$.

In order to verify the simulation results, emissive sheath theory in Equation (1) is updated considering the energy dependency of $\gamma_{e0}$. In order to derive analytical sheath solution, the abovementioned approximation of dimensionless SEE coefficient $\gamma_{e0}(\varepsilon) = \frac{\varepsilon}{\varepsilon_1}$ is used. The effective SEE coefficient for plasma electron is calculated by:

$$\gamma_{e0} = \frac{\int_0^\infty \frac{0.5 m_e v_e^2}{\varepsilon_1} v_e f_{pe,wall} dv}{\int_0^\infty v_e f_{pe,wall} dv} = \frac{T_e}{\varepsilon_1} \tag{13}$$

Hence the sheath potential becomes:

$$e\varphi_{sh} = T_e \ln[(1 - \frac{T_e}{\varepsilon_1})\sqrt{\frac{\mu}{2\pi}}] \tag{14}$$

Equation (14) is only valid for non-SCL sheath, which requires SEE coefficient smaller than the critical value $\gamma_{e0} \leq \gamma_{crit} \approx 1 - 8.3\mu^{-0.5}$ (5). Note that in the present manuscript, all temperatures, $\varepsilon_1$ and $e\varphi_{sh}$ have the unit eV to facilitate calculation. The critical emission coefficient corresponds to a marginal sheath solution featuring zero electric field at the wall, with a critical sheath potential $e\varphi_{sh,crit} \approx 1.02 T_e$. Above analyses indicate that the plasma temperature should be lower than a critical temperature $T_{e,crit}$ to stay in classic Debye sheath:

$$T_e \leq T_{e,crit} = \varepsilon_1(1 - 8.3\mu^{-0.5}) \tag{15}$$

In Figure 5(a), (b), the sheath remains in classic sheath regime for the selected temperature range. The theory prediction is consistent with simulation results applied with the third method introduced above. Simulations with method 1 and 2 are close but in general underestimate $\gamma_{e0}$. Equation (14) suggests that $\varphi_{sh}$ increases slower and even decrease as $T_e$ increases, as opposed to the classic emissive sheath theory which predicts a linear relation between sheath potential $\varphi_{sh}$ and plasma electron temperature $T_e$. This trend is due to greater induced $\gamma_e$ with electron temperature increases, which is more obvious in Figure 5(c), (d) where $\varphi_{sh}$ drops sharply to $\varphi_{sh,crit}$ at approximately 30eV. Note that above expressions for $\gamma_{e0}$ are not valid when the sheath enters the SCL mode, as the formation of local virtual cathode can reflect the emitted secondary electrons and also affect plasma electrons. Analytic formula of $\gamma_{e0}$ is difficult and is not given here, but should not influence the sheath potential as it is no longer sensitive to $\gamma_{e0}$ above the critical emission yield. This is marked by a dashed line in Figure 5(c), also the linear dependence of sheath potential on $T_e$ is restored above 30eV. Above trend is valid only for simulation with method 3 as the other two methods underestimate $\gamma_e$ so that $\gamma_{crit}$ is not achieved. The critical electron temperature is $T_{e,crit} = 45.2eV$ for $\varepsilon_{max} = 300eV$, $\gamma_{max} = 3$, and is 29.8eV for $\varepsilon_{max} = 200eV$. This is because a smaller $\varepsilon_{max}$ or larger $\gamma_{max}$ increases the slope of the left part of SEE coefficient curve, which decreases $\varepsilon_1$.

Note that the adopted kinetic sheath theories above are not valid when sheath collisionality becomes significantly higher, particularly when ion-neutral collision mean free path is well below the Debye length. Ions are limited by their mobility at high pressure levels and a clear presheath region cannot be defined. Fluid approaches are more favorable for the highly-collisional sheath, but will not be further developed in the present work.

In addition, it must be pointed out that the good agreement between kinetic sheath theory and the kinetic simulation results is partially due to the special choice of collision operator in the simulation, which ensures that the bulk plasma always follows the Maxwellian distribution of given temperature. Larger discrepancies with the theories are expected for PIC simulation with more self-consistent treatment for collisions.

## 3.3 Combined SEE model and analyses of sheath stability

In sections 3.1 and 3.2, the influences of wall charging and SEE coefficient energy dependency on sheath properties are discussed separately. Below the simulation results combining both factors are presented, applying method 3 of $\gamma_{e0}$ calculation introduced in section 3.2. The surface emission flux is calculated as:

$$\Gamma_{em} = \int_0^\infty \frac{\gamma_{e0}(0.5 m_e v_e^2)}{1 + K_{wall}\sigma_{wall}} v_e f_{pe,wall} dv \qquad (16)$$

with $\sigma_{wall} \leq 0$ and $f_{pe,wall}$ the plasma electron VDF at the wall. Simulation results are shown in Figure 6. The theory prediction of SEE coefficient and sheath potential are calculated from Equation (10), (13), (14).

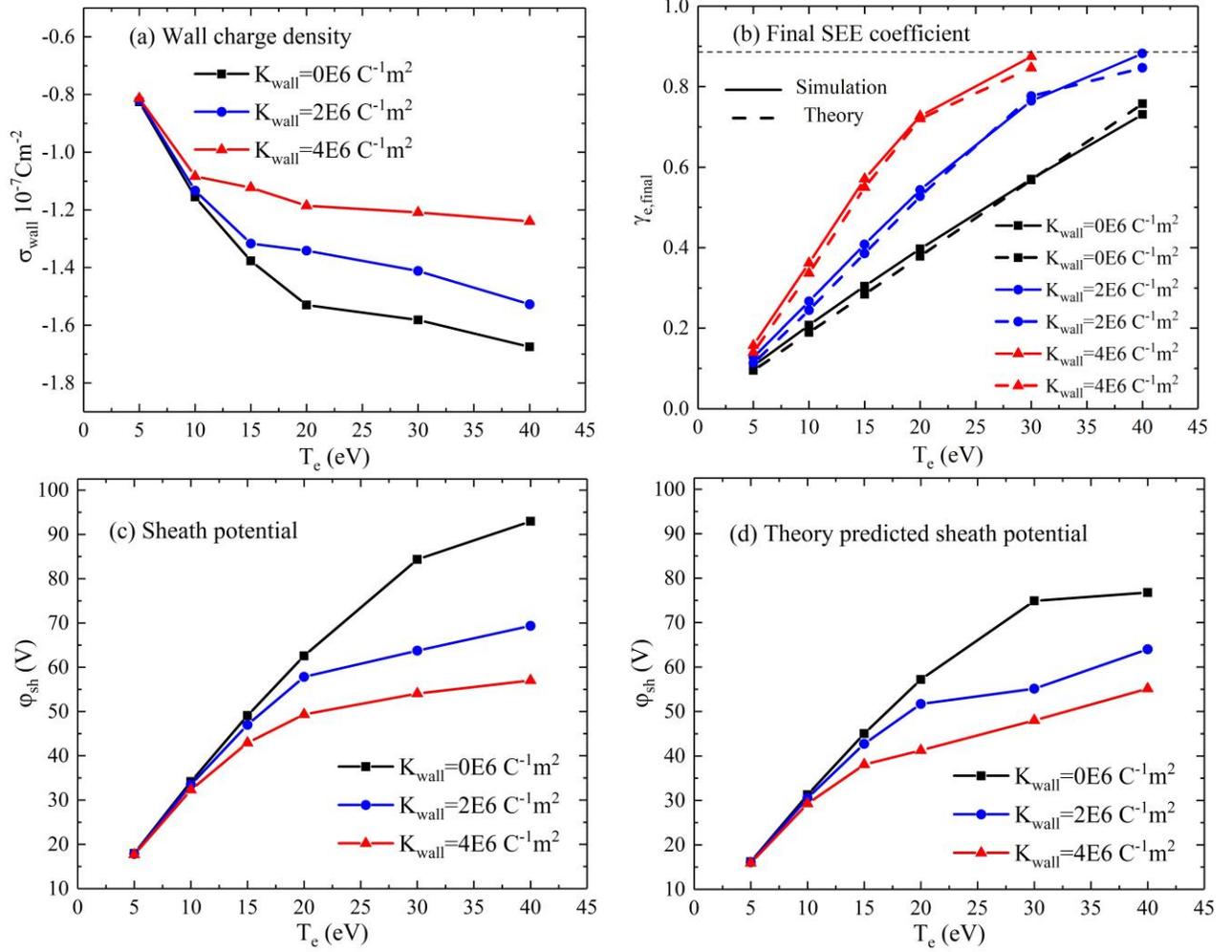

Figure 6. Emissive sheath properties with combined SEE model, for a range of $T_e$ with different wall charging factors. (a) Wall charge density. (b) Final SEE coefficient after reaching convergence, solid lines show simulation results and dashed lines show theory prediction. (c) Sheath potential from simulation. (d) Sheath potential predicted by emissive sheath theory.

The amount of wall charges increases with plasma electron temperature and decreases with the wall charging factor $K_{wall}$, see Figure 6(a). Since both higher $T_e$ and $K_{wall}$ lead to higher SEE coefficient, $\gamma_e$ achieves the critical value marked by dashed lines at $K_{wall} = 4E6 C^{-1}m^2$ and $T_e \geq 30 eV$, see Figure 6(b). Sheath potential given by simulation and updated sheath theory show consistent trend with respect to $T_e$, which increases slower at higher $T_e$ levels due to energy dependency of $\gamma_e$, and decreases with $K_{wall}$ due to wall charging.

One critical implication from above analyses is, whether the inclusion of charging effects and $\gamma_e$'s energy dependency will affect the I-V characteristic of the sheath. I-V characteristic of emissive sheath is closely linked to the sheath stability and hence practical plasma applications. The sheath stability analyses considering both $\gamma_e$'s energy dependency and charging effects is analyzed as follows, which are based on previous emissive sheath stability study assuming constant SEE coefficient (33). The net electron flux to the dielectric wall is $\Gamma_e = \Gamma_{ep} - \Gamma_{em} = (1-\gamma_e)\Gamma_{ep}$. Here $\gamma_e$ depends on incident electron energy and the wall charge density. To guarantee a stable emissive sheath, the derivative $\frac{\partial \Gamma_e}{\partial \varphi_{sh}}$ should be negative. A perturbation, e.g. small increase of $\varphi_{sh}$, indicating that more negative charges are accumulated in the wall, must cause the net electron flux to decrease so that the sheath is restored back to equilibrium. The derivative $\frac{\partial \Gamma_e}{\partial \varphi_{sh}}$ is further developed as:

$$\frac{\partial \Gamma_e}{\partial \varphi_{sh}} = (1-\gamma_e)\frac{\partial \Gamma_{ep}}{\partial \varphi_{sh}} - \frac{\partial \gamma_e}{\partial \varphi_{sh}}\Gamma_{ep} < 0 \tag{17}$$

Here the sheath is assumed to be classic Debye sheath with $\gamma_e < 1$. The wall plasma electron flux is $\Gamma_{ep} = \Gamma_{ep,se}\exp(-\frac{e\varphi_{sh}}{T_{ep}})$ with $\Gamma_{ep,se}$ the plasma electron flux at sheath edge. The first term in RHS of Equation (17) is hence negative. The derivative $\frac{\partial \gamma_e}{\partial \varphi_{sh}}$ is further expressed as follows combining Equations (10), (13):

$$\frac{\partial \gamma_{e0}}{\partial \varphi_{sh}} = \frac{1}{1+K_{wall}\sigma_{wall}}\frac{1}{\varepsilon_1}\frac{\partial T_e}{\partial \varphi_{sh}} - \frac{K_{wall}}{(1+K_{wall}\sigma_{wall})^2}\frac{T_e}{\varepsilon_1}\frac{\partial \sigma_{wall}}{\partial \varphi_{sh}} \tag{18}$$

Note that $\sigma_{wall} < 0$ in the present analyses. If a constant plasma electron temperature is assumed, which is true for the present simulation but may vary in practical discharges, the RHS of Equation (18) should be positive since $\frac{\partial \sigma_{wall}}{\partial \varphi_{sh}} < 0$. This is because the amount of negative charges trapped in wall is equal to the positive charge amount in sheath, due to charge conservation of a floating wall. Similar trend is also shown in Figure 3(a)(c) where $\sigma_{wall}$ and $\varphi_{sh}$ change oppositely to $\gamma_e$. As a result, considering charging effect does not alter the emissive sheath stability since the RHS of Equation (17) remains strictly negative. The conclusion is reassuring for applications such as plasma processing where stable plasma flux is expected.

3.4 Implementation of electron backscattering in simulation

In above sections, the algorithms to implement secondary electron emission in the kinetic simulation are discussed in detail. Apart from SEE, some incident electrons on solid wall are reflected back after elastic interactions with the sample atoms. Such process is called backscattering and in general occurs in a deeper region than SEE. The most distinct difference between secondary electron and backscattered electron is their velocity distribution functions. Backscattered electrons velocity scales up with plasma electrons, whereas SEs mostly have low energy regardless of incident electron velocity.

A simple way to implement backscattering is to discard the different physical mechanisms of SEE and backscattering, assuming that both types of electron have the same temperature and using an effective emission coefficient as follows:(20)

$$\gamma_{eff} = (1-R_b)\gamma_e + R_b \tag{19}$$

with $R_b$ the backscattering probability. The treatment simplifies the backscattering process as a special SEE with $\gamma_e = 1$, whose implementation in both analytical analyses and simulation is straightforward by replacing $\gamma_e$ with $\gamma_{eff}$. The assumption adopted by Equation (19) is inevitably less accurate for high temperature electrons due to fast backscattered electrons. For kinetic simulation, a more physical treatment is to use the following boundary condition for the EVDF:

$$f_e(v_e)|_{v_e>0, x=0} = (1 - R_b)f_{em} + f_{eb} \tag{20}$$

Here $f_{eb}$ is the backscattered electron VDF. The separation of secondary electron and backscattered electron VDF is due to their different temperatures. Temperature of secondary electron is fixed for given wall material, while plasma electron temperature in practical discharge varies depending on particle and energy balances. If zero energy loss is assumed in the backscattering process, the reflected electron VDF is proportional to $f_{eb} \propto \exp\left(-\frac{m_e v_e^2}{2T_e}\right)$. Using the definition of reflection coefficient $\Gamma_{ep} R_b = \Gamma_b$ and backscattering flux $\Gamma_b = \int_0^\infty f_{eb} v_e dv_e$, $f_{eb}$ is calculated to be:

$$f_{eb} = \frac{m_e}{T_e} R_b \Gamma_{ep} \exp(-\frac{m_e v_e^2}{2T_e}) \tag{21}$$

For the present simulation with Maxwellian plasma electron, Equation (20) is equivalent to:

$$f_e(v_e)|_{v_e>0, x=0} = (1 - R_b)f_{em} + R_b f_e(v_e)|_{v_e<0, x=0} \tag{22}$$

Here the left wall boundary is taken as an example, where the $v_e < 0$ part of $f_e$ points to the left wall. In the adopted code, above equation is implemented by setting $v_e > 0$ part of $f_e$ at left boundary to the sum of $(1 - R_b)\frac{m_e}{T_{em}} \gamma_e \Gamma_{ep} \exp(-\frac{m_e v_e^2}{2T_{em}})$ and $R_b$ times the $v_e < 0$ part of $f_e$. $\Gamma_{ep}$ is obtained by integrating $v_e$ over $f_e$ (only for $v_e < 0$), $\gamma_e$ and $R_b$ are given as constant. The right wall boundary condition should be symmetrical.

The EVDFs at the left wall with two different boundary conditions mentioned above are shown in Figure 7. It is clear that the boundary EVDFs using two above methods are remarkably different mainly due to different emitted electron temperatures. Since the SEE-emitted electron temperature is typically within 5eV whereas backscattered electrons have the same temperature as the plasma electrons, the discrepancy is obvious only for simulations with high $T_e$. The first method (Equation (19)) usually yields a $f_{eb}$ more centralized than the 2nd method (Equation (20)).

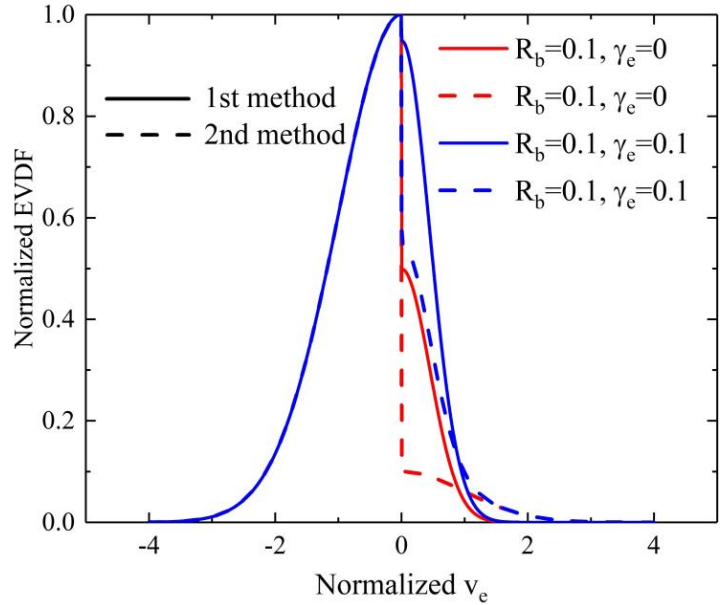

Figure 7. EVDF at the left wall with two different methods to set boundary condition with backscattering.

The difference in boundary EVDF, however, is shown to have limited influences on simulation results, shown in Figure 8. A scan of true SEE coefficient with different backscattering coefficients are tested, with the effective SEE coefficient calculated as $\gamma_{eff} = \Gamma_{em}/\Gamma_{ep}$. Here $\Gamma_{em}$ contains contributions from both SEE and reflection. Note that for low energy primary electrons, the reflection rate is unlikely to be very high and is typically below 0.2 (34). The effective SEE coefficient calculated from simulation using Equation (20) is only slightly greater than the constant effective SEE coefficient prescribed by Equation (19). The sheath potentials are also almost equal accordingly. The conclusion is interesting as it suggests that the emissive sheath properties are not affected by using secondary electron to replace backscattered electrons via a converting relation dictated by Equation (19), assuming incoming plasma electron flux is the same. Though having completely different physical mechanism, the two types of electrons have similar contribution to the emissive plasma sheath properties. Adopting such assumption will greatly simplify related theoretical works.

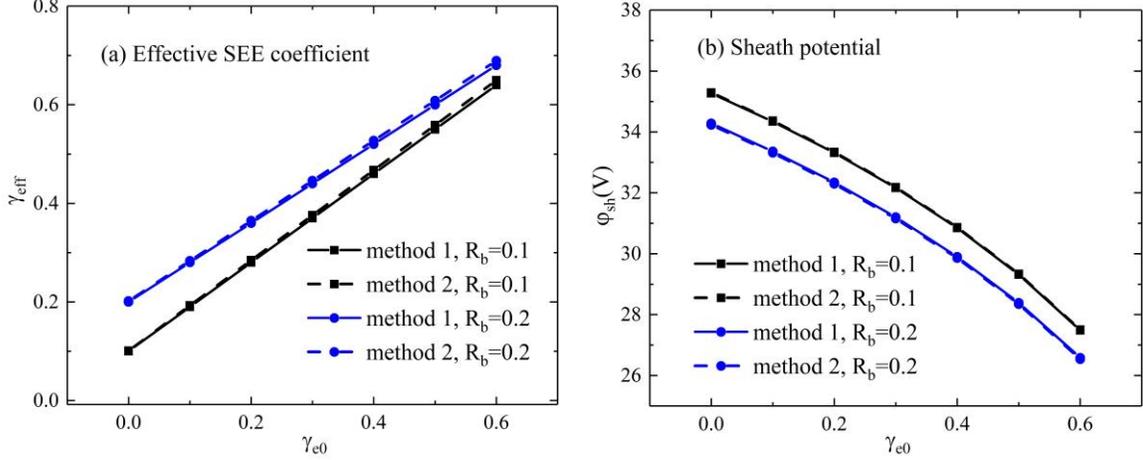

Figure 8. Effective SEE coefficient and sheath potential with a scan of true SEE coefficient and different backscattering coefficients.

With above conclusions, the emissive sheath potential with both SEE and backscattering is simply:

$$e\varphi_{sh} = T_e \ln[(1-(1-R_b)\gamma_e - R_b)\sqrt{\frac{m_i}{2\pi m_e}}] \tag{23}$$

Note that conclusion above is valid for purely elastic backscattering. If a constant fraction of energy $\eta = T_{eb}/T_{ep}$ is lost after the backscattering with $\eta$ the dimensionless backscattering energy loss factor, and $T_{eb}$ the backscattered electron temperature, the predicted effective emission coefficient using Equation (19) is not affected, whereas the EVDF boundary condition in Equation (20) should be replaced by:

$$f_e(v_e)|_{v_e>0,x=0} = (1-R_b)f_{em} + \frac{m_e}{\eta T_e} R_b \Gamma_{ep} \exp(-\frac{m_e v_e^2}{2\eta T_e}) \tag{24}$$

The influence of energy loss factor $\eta$ is shown in Figure 9. The factor $\eta$ is found to have negligible influence on the sheath potential and effective SEE coefficient. The limited influence of emitted electron temperature on sheath property is consistent with the emissive sheath theory, which predicts that the sheath potential depends only on the plasma electron temperature and effective SEE coefficient, instead of emitted electron temperature. More recent kinetic theory considering truncated plasma electron VDF suggested that the influence of the ratio of plasma and emitted electron temperature $\Theta = T_e/T_{em}$ does affect the sheath property but the effect is obvious only when $T_e$ is close to $T_{em}$ (10), which is unlikely to be achieved here as the backscattered electron coefficient is low and effective temperature of all emitted electrons is close to $T_{em}$.

4Running Title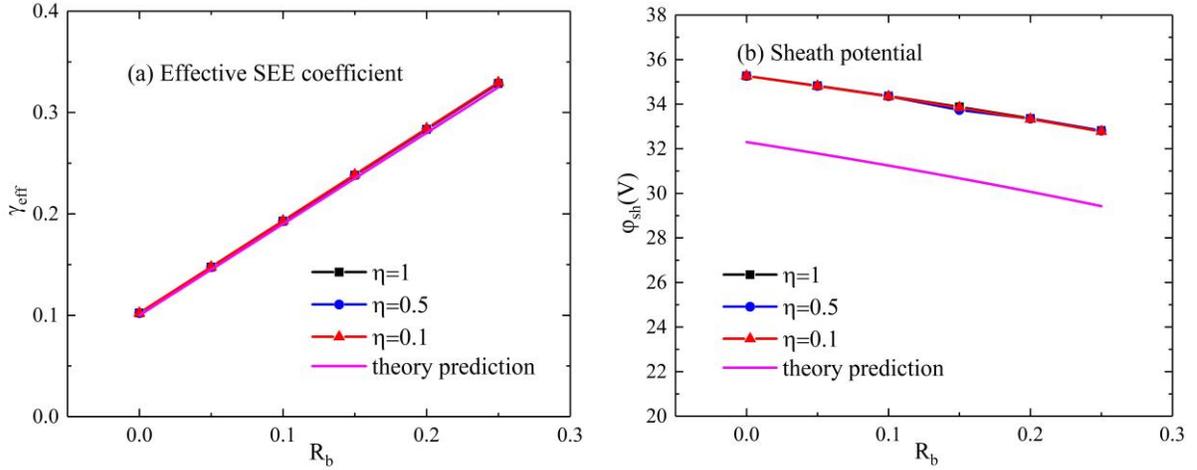

Figure 9. Effective SEE coefficient and sheath potential with a scan of backscattering coefficient using different η. The true SEE coefficient is kept as 0.1 and the theory prediction is given by Equation (23).

As has been pointed out in section 3.2, the use of artificial collision operator, though providing good agreement with the emissive sheath theory, can conceal certain physics that are only available in e.g. PIC model where self-consistent plasma-neutral collisions are implemented. A comparison of the kinetic simulation and PIC simulation was recently performed for the capacitively-coupled plasma subject to strong electron emission from the boundary (13, 35), where discrepancies are more obvious at high neutral pressure levels as the kinetic model does not implement realistic electron-neutral collisions and ionization sources. Similar comparison between kinetic and PIC model of the SEE coefficient charging effect are expected for DC or RF plasma conditions.

## 4    Conclusions

One-dimensional kinetic simulation of plasma sheath is performed involving secondary electron emission and electron backscattering. The influence of accumulated wall charges on SEE coefficient is considered and is shown to enhance the surface electron emission and decrease the sheath potential. Using energy-dependent instead of static SEE coefficient is found to induce nonlinear sheath potential response to the plasma electron temperature, as opposed to the classic emissive sheath theory. SCL sheath is formed if plasma electron temperature or wall charge density is sufficiently high so that effective SEE coefficient is above a critical value. The EVDF boundary condition for electron backscattering is proposed and implemented in the kinetic simulation. Considering backscattering electron mainly affects boundary EVDF as backscattered electrons are typically faster than secondary electrons. Converting backscattered rate into SEE coefficient via an equivalent equation is shown to barely affect the sheath properties. The simulation results are well supported by the upgraded emissive sheath theories, where wall charging effect, SEE coefficient's energy dependency, and backscattering are included.

## 5    Author Contributions

GS, BG, and SZ contribute to the concept, modeling and theory, AS and GZ supervise the work throughout. All authors contributed to the final version of the manuscript.

## 6    Funding

The research was conducted under the auspices of the National Key R&D Program of China (No. 2020YFC2201100) and National Natural Science Foundation of China (Nos. 51827809, 52077169). This work was also supported in part by the Swiss National Science Foundation.

**Data Availability Statement**

The data that support the findings of this study are available from the corresponding author upon reasonable request.